\documentclass{sig-alternate}
\usepackage{url}
\usepackage{graphicx}
\usepackage{amsmath}

\begin{document}
\conferenceinfo{VSW06}{June, 2006, Berlin, Germany}
\CopyrightYear{2006} 
\crdata{1-59593-387-5}
\title{Intrusion detection mechanisms for VoIP applications}

\numberofauthors{3}
\author{
\alignauthor Mohamed Nassar \\
\affaddr{LORIA - INRIA Lorraine} \\
\affaddr{615, rue du jardin botanique, 54602, Villers-L\`es-Nancy, France} \\
\email{nassar@loria.fr} 
\alignauthor Radu State \\
\affaddr{LORIA - INRIA Lorraine} \\
\affaddr{615, rue du jardin botanique, 54602, Villers-L\`es-Nancy, France} \\
\email{state@loria.fr} 
\alignauthor Olivier Festor \\
\affaddr{LORIA - INRIA Lorraine} \\
\affaddr{615, rue du jardin botanique, 54602, Villers-L\`es-Nancy, France} \\
\email{festor@loria.fr}
}
\maketitle
\begin{abstract}
VoIP applications are emerging today as an important component in business and communication industry. In this paper, we address the intrusion detection and prevention in VoIP networks and describe how a conceptual solution based on the Bayes inference approach can be used to reinforce the existent security mechanisms. Our approach is based on network monitoring and analyzing of the VoIP-specific traffic. We give a detailed example on attack detection using the SIP signaling protocol.
\end{abstract}
\section{Introduction}
With VoIP we inherit the adjacent security problems associated with the IP as well as new VoIP specific ones. Attackers can profit from the vulnerabilities of the VoIP protocols and architectures. Both Signaling protocols such as H.323 and SIP (Session Initiation Protocol), and media transport protocols such as RTP and RTCP could be the target of a wide set of attacks, ranging from eavesdropping, denial of service, fraudulent usage and SPIT (Spam over internet telephony).
 
Important work in both host and network intrusion detection has been already done by the industrial and academic research community, focused in scope towards network intrusion detection for routing, transport and application level protocols. However, specific approaches for VoIP are still in an incipient stage and we were motivated in our work to leverage existing conceptual solutions for the VoIP specific application domain. Our paper is structured as follows: a brief review of the possible VoIP specific threats is given in section \ref{attacks}. An introduction of the Bayesian inference theory is given in the section \ref{bayesinf}. Section \ref{bayesmodel} will present a detailed approach of a network-based intrusion detection using a statistical Bayes model. Section \ref{related} overviews the related works and section \ref{conclusion} concludes the paper.

\section{VoIP threats}
\label{attacks}
Although, the SIP protocol has some security capabilities (like for instance to partially encrypt messages) and thus making eavesdropping and media spamming harder, some other threats represent real sources for major damages. The encryption of message headers is allowed, but headers like \texttt{To, From, Call-ID, CSeq} and \texttt{Via} are important to the proxies, such that its usage is limited. TLS can be used for a hop by hop based encryption and end to end encryption,  non repudiation and integrity are possible via S/MIME, but performance is heavily impacted. In the following paragraphs, we go through a panorama of the most important threats.

\paragraph*{Messages interception and call tracking}
The SIP \texttt{INVITE} packets contain sensible informations such as the source and the destination of the call (\texttt{To},  \texttt{From}, \texttt{Via} headers), allowing call tracking. The duration of a call can be calculated by logging the time of the \texttt{INVITE} message who started the call and the \texttt{BYE} message who ended it. Interception of unencrypted SIP requests and handling of its values can be a preparatory stage to session hijacking and man in the middle attacks. 

\paragraph*{Fraudlent usage}
Malicious clients could try to bypass the billing procedure and to make calls to the PSTN (Public Switched Telephone Network) for free.  Unsecured gateways, misconfigured dialplans and platform specific vulnerabilities are common causes for it. 

\paragraph*{Password cracking and user enumerating}
Some attacks attempt to break down the access control, like for instance brute force  password cracking. These attacks could be preceded by a scanning and enumeration of existing user names. A scanner which is looking to know the valid numbers in a domain will send a sequence of requests investigating the location server of the domain. Each request carries a different number or user name as destination. The scanner will bind between the requests and the corresponding responses to filter up the existent destinations.

\begin{figure*}
	\centering
	\includegraphics[width = 400pt, height = 200pt]{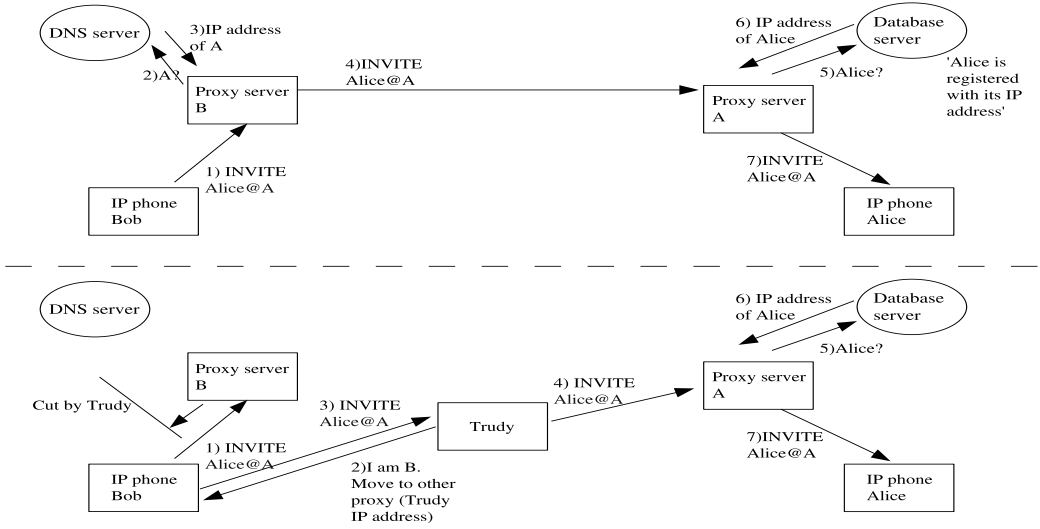}
	\caption{Normal and Attack scenarios}
	\label{fig:hijack}
\end{figure*}

\paragraph*{Call hijacking and man in the middle attack}
SIP uses a strong authentication scheme similar to the HTTP digest. A SIP user agent, proxy, redirect or registrar server might challenge a client user agent to authenticate. In addition, a user agent can challenge back a proxy server to be sure that it also knows the shared secret. Unfortunately, only the server authenticates the client in most VoIP implementation and the following attack is possible:
 Bob wants to call Alice. Bob's phone sends an \texttt{INVITE} message to the proxy server within its domain in order to route the call towards Alice's domain. Trudy impersonates the proxy and redirects the call to its own IP address. To do so, Trudy has to create a crafted SIP response and put its own IP address in the \texttt{Contact} header. Bob's phone does not require the authentication of the proxy, so it accepts the response and redirects the call to pass by Trudy's machine. In figure \ref{fig:hijack}, a normal call setup is shown at the upper side and a scenario of call hijacking is shown at the lower side. We do not show all the SIP messages involved in both scenarios but just the main steps numbered by order of appearance. 

If the end to end authentication is missed -which is a common case-, man in the middle scenario is possible. A session established between Bob and Alice could be hijacked by Trudy. Indeed, Trudy who has caught the \texttt{INVITE} in the origin of the session, copies the \texttt{Call-ID, To }and \texttt{From} tags to a new \texttt{INVITE} packet and increments the sequence number \texttt{CSeq}. Trudy steps in and sends the crafted \texttt{INVITE} to Bob or Alice. Since no authentication is required, Trudy ends up by hijacking the session. She can change both media and signaling characteristics, like for example changing RTP ports, adding or deleting media streams, changing the signaling path (\texttt{Via} headers) or denying signaling from any side to its benefit (\texttt{Contact} header).

\paragraph*{Denial Of Service (DOS)}
This type of attacks aims to affect the availability of the service. Attackers can search for vulnerabilities in the VoIP stack of a server to find a way to take it down using malformed packets. Otherwise, they can fill up the available bandwidth by flooding traffic so the server could not use the network ressources. 

CANCEL and BYE attacks might take place against the SIP call establishment procedure. If whenever someone tries to call Bob, Trudy sends a \texttt{CANCEL} to Bob, then Bob will be prevented from receiving calls. If whenever Bob tries to make a call, Trudy sends a \texttt{CANCEL} to the destination, then Bob will be prevented from making calls. Otherwise, Trudy could send a \texttt{BYE} to terminate a session after a few moments of its setup. Trudy could use proxy responses such as 4xx(client error), 5xx(server error) or 6xx(global error) to convince Bob that an error situation is preventing him from making calls.

A traditional DOS can be launched against a stateful SIP proxy or a gateway in form of a large amount of requests with different \texttt{Call-Ids}. Distributed sources could participate in the attack in order of surcharging the server capacities (memory, CPU or bandwidth). A list of VoIP specific DOS could be found in \cite{voipsa:tax}. 

\paragraph*{Attacks against gateways and voice mail servers}
The gateway between PSTN and VoIP networks is a critical point from the billing perspective. Most often, It is the host where the Call Detail Records (CDRs) are safeguarded and the accounting operations are proceeded. The voice mail servers may contain confidential informations about the customers. For these reasons, the gateways and the mail servers are the probable target for the hackers activities. These people will try different kinds of host intrusions or remote code execution and buffer overflow attacks aiming to hijack the configuration or to tamper with the data. 

\paragraph*{SPIT or SPAM over Internet telephony}
The SPAM unwanted messages that threaten the e-mail users will be joined by a more annoying voice advertising. The nature of the Internet protocols gives easy ways to stream real-time voice messages to a large number of destinations. When they are filled by automated calls, IP phones and voice-mail boxes will be useless.

\paragraph*{Media protocols related attacks}
Modifications of the media characteristics by a media protocol is not transparent for SIP. Actually, SIP could be classed as an out-band signaling and does not have control mechanisms to sense a changing in the media session. In addition, multimedia protocols such as RTP and RTCP have their own vulnerabilities. A demonstrative tool of the RTP play out attack was presented in \cite{nmrg}. The encryption of the RTP streams seems to be a good solution to prevent eavesdropping, but the attackers may have got the encryption keys by rather than a way (for instance if the keys are negotiated clearly in the media negotiation phase). 

\paragraph*{Supporting protocols related attacks}
The ARP (Address Resolution Protocol) poisoning attack consist in binding the physical address of the intruder with the IP address of the gateway at the IP phone machine, and on binding the physical address of the intruder with the IP address of the IP phone at the gateway machine. The DNS (Domain Name System) poisoning attack can be also used to perform man in the middle attacks. MAC and IP spoofing are fundamental flaws in the basic Internet and could not addressed from a particular application point of view. 

\paragraph*{Firewall traversal}
The firewalls utilized with VoIP have dynamic skills. They open and close RTP ports with respect to the SDP (Session Description Protocol) parameters during the session initiation. The firewall could be attacked by making it dealing with a large number of port opening requests so it may loose its defense functionality.

\section{Introduction to  the Bayes inference}
\label{bayesinf}
Bayesian methods provide a formalism for reasoning about partial belief under conditions of uncertainty \cite{pearl}. They are based on the empirically verifiable relationship between posterior(the belief we accord a hypothesis H upon obtaining evidence e) and prior(P(H)) probabilities: 
$$ P(H/e) = \frac {P(e/H) P(H)}{P(e)}$$
A Bayesian network is a directed acyclic graph whose arrows represent causal influences and each of its nodes represents certain knowledge and is considered to be in one of several discrete states. In a Bayesian tree, each node might have several children and one parent. The propagation and fusion of the belief in a Bayesian tree are proceeded under the following rules:
\begin{itemize} 
\item The likelihood (or diagnostic) messages $\lambda$ are travelling upward the tree. 
\item The prior (or causal) messages $\pi$ are travelling downward the tree. 
\item A child is linked to its parent by a conditional probability table (CPT) of which the elements are given by: $$CPT_{ij}=P(child=j/parent= i)$$ Each row of the matrix is a discrete distribution over the child node states giving the parent node state and thus it sums to 1. 
\item The propagation of the prior messages is given by: $$\pi (child)= \alpha \pi (parent) \bullet CPT(child/parent)$$ where $\pi$ is a row vector and $\alpha$ is a constant to normalize the distribution. 
\item The propagation of the likelihood messages is given by: 
\footnotesize
$$\lambda_{to\_parent} (child) = CPT(child/parent) \bullet \lambda (child)$$  
\normalsize
where $\lambda$ is a column vector.
\item The likelihood messages are fused together by an elementwise multiplication: 
\footnotesize
$$ L_{i}(parent) = \prod_{child \in children(parent)} \lambda_{{to\_ parent}_i}(child) $$ 
\normalsize
$\lambda (parent)$ \ is \  obtained  \ by \ normalizing \ the \  vector \ $L(parent)$ to the unit sum. 
\item Finally, the belief over the states at a node is obtained by an elementwise multiplication of $\lambda (parent)$ and $\pi (parent)$ and then normalizing the resulting vector by an appropriate constant $\beta$: $$BEL_i=\beta \pi_i \lambda_i$$
\end{itemize}
\section{Bayes  model  to  detect  intrusive SIP  traffic}
\label{bayesmodel}
In this section, we present the exercise of applying the Bayes tree model into a network-based intrusion detection solution for VoIP. The model was firstly applied in the intrusion detection domain as a component of the broad EMERALD system to detect intrusive TCP sessions \cite{alf:adaptive}. The source of data for our detector engine is captured SIP traffic. The packets of other VoIP protocols such as RTP or VoIP supporting (DNS) could be also an important source of data which could be exploited in future works.
\subsection{The model structure}
In our context, the source of an incoming request is the value of the last \texttt{Via} header of this request, because the response will be routed back to arrive to this address. The destination of a request can be found in the first line of the request. \texttt{From} and \texttt{To} headers are logical source and destination. In the following example, \texttt{here.com} is the SIP source of the \texttt{ACK} request, and \texttt{UserB@there.com} is the destination.
\footnotesize
\begin{verbatim}
ACK sip:UserB@there.com SIP/2.0
Via: SIP/2.0/UDP ss2.wcom.com:5060;branch=721e418c4.1
Via: SIP/2.0/UDP ss1.wcom.com:5060;branch=2d4790.1
Via: SIP/2.0/UDP here.com:5060
From: BigGuy <sip:UserA@here.com> 
To: LittleGuy <sip:UserB@there.com> ;tag=314159
Call-ID: 12345601@here.com
CSeq: 1 ACK
Content-Length: 0
\end{verbatim} 
\normalsize

To construct the model, choose the different variables, study the dependency relationships between variables and set the different parameters, a large empirical database is needed. However, and due to our poverty to real world traces, we have recourse to our knowledge of how SIP works and how attacks are performed to extract a first prototype that could be implemented to defend a VoIP site and updated increasingly while new experiences are acquired by the time.

We use a na\"ive Bayes model which is a 2-level tree formed by one root node and several leaf nodes. The root represents the traffic class which is unobservable. The leafs represent the directly observable evidences. We assume that the child nodes are conditionally independent given the parent: $$P(child1/parent) = P(child1/child2,parent)$$ $\forall \  child1, child2 \in children(parent)$.

The belief propagation and updating is done periodically and we call it the inference process. The period could be configured as a count of occurred events number or as a measure of elapsed time. The events are either a SIP message captured or an exception thrown in deciphering or parsing. During the period time, the events update the values of variables at the leaf nodes so at the end of each period the likelihood messages could be estimated. While the most important in the interpretation at the class traffic root is to distinguish between attack and non attack cases, our model includes the following states of interest: \textsf{NORMAL, DOS, SCAN, PASSWORD CRACKING, FIREWALL TRAVERSAL} and \textsf{SPIT}.

The observed variables at the leaf nodes are of three types: \textsf{Request Intensity(RI), Error Response Intensity(ERI)} and \textsf{Parsing Error Intensity(PEI)} are intensity measures, \textsf{Number of Different Destinations, Max Number of Dialogs in Waiting State} and \textsf{Number of opened RTP ports} are high water marks, finally \textsf{Request Distribution} and \textsf{Response Distribution} are distribution measures.
Intensity measures are exponentially decayed counts. Noting that the codes of the SIP error responses are between 300 and 699, let $I(resp\_code) = 1$ if $300 \leq resp\_code$ and $I(resp\_code) = 0$ otherwise. Intensity measures are computed by the following formulas:
\begin{itemize}
\footnotesize
\item $RI_{req} =  e^{k\Delta t} \cdot RI_{req-1} + 1.0 \; \text{;}$
\item $ERI_{resp} = e^{k \Delta t} \cdot ERI_{resp-1} + I(resp\_code);$
\item $PEI_{err} = e^{k \Delta t}\cdot PEI_{err-1} + 1.0 \; \text{;} $
\normalsize
\end{itemize}
Where $\Delta t$ is the time between the present and the immediately preceding event and $k$ is a decay constant and is $\le 0$.
The most appropriate is to measure the time chronologically in case of \textsf{RI} and by events count in case of \textsf{ERI} and \textsf{PRI}. Like that, An exponentially decayed count does not grow without a bound for normal behavior and well chosen decay constants. We divide the intensity range into several intervals. At each end of period, the intensity measure falls under one interval. So, the likelihood probability is 1 for the matched interval and 0 for all other ones.

We initiate counters to record the high water marks measures. The \textsf{max number of dialog in waiting state} is incremented for each request which is responded but it holds the state machine of the dialog waiting for an \texttt{ACK}. Once time the \texttt{ACK} is received, the counter is decremented by one. The maximum reached by the counter at any time is tracked. Like the intensity measures, we divide the high water mark range into several intervals and we assign appropriately the likelihood probabilities at the end of a period. After the inference process is finished, we reset all the counters.

\begin{figure*}
	\centering
	\includegraphics[width = 300pt, height = 200pt]{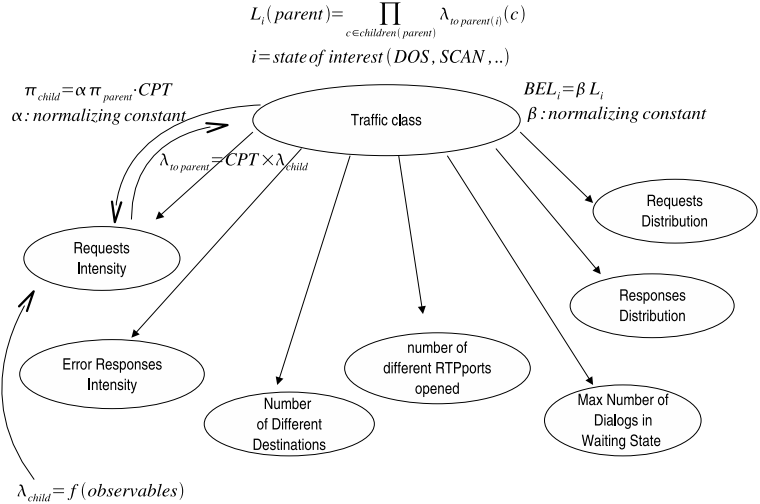}
	\caption{Bayes model for SIP}
	\label{fig:sipbayes}
\end{figure*}

Between 2 inference processes, we track the count of each request to build up the \textsf{Request Distribution}. Likewise, the responses are grouped by their respective classes and counted to build up the \textsf{Response Distribution}. The distribution is taken as the likelihood vector of the leaf node.
The figure (\ref{fig:sipbayes}) depicts the scheme of our model.

\subsection{Example of the attack detection process}
We aim in this section to go through the inference process starting from a trace of attack to clarify and motivate our approach. The calculation is not totally complete because we have no empirical idea about the prior probabilities of the different traffic classes so the propagation will be in only one direction rather than two (upward).
 
The following trace is an example of a URI scanning (enumerating) attack in which the attacker tries to call 9 different SIP URIs behind a user agent serving multiple clients. We collect for each dialog the incoming requests from the attacker and the  outgoing responses from the user agent.
\footnotesize
\begin{flushleft}
Dialog 1: \texttt{INVITE $\rightarrow$ 404 Not Found $\rightarrow$ ACK}\\
Dialog 2: \texttt{INVITE $\rightarrow$ 484 Address Incomplete $\rightarrow$ ACK}\\
Dialog 3: \texttt{INVITE $\rightarrow$ 100 Trying $\rightarrow$ 503 Service Unavailable $\rightarrow$ ACK}\\
Dialog 4: \texttt{INVITE $\rightarrow$ 100 Trying $\rightarrow$ 180 Ringing $\rightarrow$ CANCEL $\rightarrow$ 200 0K(CANCEL) $\rightarrow$ 487 Request Terminated $\rightarrow$ ACK} \textit{Good number, the attacker hangs up immediately.}\\
Dialog 5: \texttt{INVITE $\rightarrow$  404 Not Found $\rightarrow$ ACK}\\
Dialog 6: \texttt{INVITE $\rightarrow$ 484 Address Incomplete $\rightarrow$ ACK}\\
Dialog 7: \texttt{INVITE $\rightarrow$ 100 Trying $\rightarrow$ 503 Service Unavailable $\rightarrow$ ACK} \textit{The number could be right but his owner is not registered at the moment}\\
Dialog 8: \texttt{INVITE $\rightarrow$ 100 Trying $\rightarrow$ 180 Ringing $\rightarrow$ 200 OK $\rightarrow$ ACK $\rightarrow$ BYE $\rightarrow$ 200 OK} \textit{Good number, the call is answered, the attacker hangs up.}\\
Dialog 9: \texttt{INVITE $\rightarrow$ 404 Not Found $\rightarrow$  ACK}\\
\end{flushleft}
\normalsize
Let us first set up the CPT matrices that relates between the root and the children nodes. Theses matrices are normally evaluated by a learning phase, but here we will set them manually according to protocol semantics. For simplicity sake, we will divide the range of each measure (except distributions) into a few number of categories. More experiments with traces of attacks will allow us in the future to refine the number and the borders of categories.
We assume that no S/MIME encryption mechanism was in place, and no parsing errors have occurred. The \textsf{PCE} is always 0 so the related CPT has no influence on the inference process and it is not shown.
In the \textsf{Request Intensity} formula, we set the decay rate at $-0.35$ for a half-life time of 2 seconds. To fill up the CPT tables, we asked such kind of questions: If the traffic is of kind DOS, what is the probability to have $RI > 10$. 
\begin{center}
\footnotesize
\begin{tabular}[htbc]{|l|c|c|}
\hline
Request Intensity	&0-10	&$>10$	\\
\hline
NORMAL 			&1	&0	\\
SCAN			&1	&0	\\
SPIT 			&1	&0	\\
DOS 			&0	&1	\\
PASSWORD CRACKING	&1	&0	\\	
FIREWALL TRAVERSAL	&1	&0	\\
\hline
\end{tabular}
\end{center}
\normalsize

In the \textsf{Error Response Intensity} formula, we measure the time as events related. We set the decay rate as $-0.15$ for a half-life time near of 5 events. We set the CPT matrix to the next:
\begin{center}
\footnotesize
\begin{tabular}[htbc]{|l|c|c|}
\hline
Error Response Intensity	&0-4	&$>4$	\\
\hline
NORMAL				&1	&0	\\
SCAN				&0.2	&0.8	\\
SPIT				&0.2	&0.8	\\
DOS				&0	&1	\\
PASSWORD CRACKING		&0	&1	\\
FIREWALL TRAVERSAL		&1	&0	\\
\hline
\end{tabular}
\end{center}
\normalsize
We set the CPT matrix of the \textsf{Number of Destinations} to the next: 
\begin{center}
\footnotesize
\begin{tabular}[htbc]{|l|c|c|}
\hline
Number of Destinations	&0-7	&$>7$	\\
\hline
NORMAL			&1	&0	\\
SCAN			&0	&1	\\
SPIT			&0	&1	\\
DOS			&0.8	&0.2	\\
PASSWORD CRACKING	&1	&0	\\
FIREWALL TRAVERSAL	&0.8	&0.2	\\
\hline
\end{tabular}
\end{center}
\normalsize
We set the CPT matrix of the \textsf{Number of Opened RTP ports} to the next:
\footnotesize
\begin{center}
\begin{tabular}[htbc]{|l|c|c|}
\hline
Number of Opened RTP ports	&0-10	&$>10$	\\
\hline
NORMAL				&1	&0	\\
SCAN				&1	&0	\\
SPIT				&0.8	&0.2	\\
DOS				&0.8	&0.2	\\
PASSWORD CRACKING		&1	&0	\\
FIREWALL TRAVERSAL		&0	&1	\\
\hline
\end{tabular}	
\end{center}
\normalsize
We set the CPT matrix of the \textsf{Max of Dialogs in Waiting State} to the next:
\begin{center}
\footnotesize
\begin{tabular}[htbc]{|l|c|c|}
\hline
Max Number of Dialogs in Waiting State	&0-10	&$>10$	\\
\hline
NORMAL					&1	&0	\\
SCAN					&0.8	&0.2	\\
SPIT					&1	&0	\\
DOS					&0.1	&0.9	\\
PASSWORD CRACKING			&0.8	&0.2	\\
FIREWALL TRAVERSAL			&0.8	&0.2	\\
\hline
\end{tabular}
\end{center}
\normalsize
We restrict the \textsf{Request Distribution} to only \texttt{INVITE(I), RE\-GISTER(R), ACK(A), CANCEL(C)} and \texttt{BYE(B)} SIP methods. We set the CPT matrix to the next:
\begin{center}
\footnotesize
\begin{tabular}[htbc]{|l|c|c|c|c|c|}
\hline
Request			&I	&R	&A	&C	&B	\\
\hline
NORMAL			&0.30	&0.10	&0.30	&0.10	&0.10	\\
SCAN			&0.40	&0.05	&0.40	&0.10	&0.05	\\
SPIT			&0.40	&0.00	&0.40	&0.00	&0.20	\\
DOS			&0.90	&0.10	&0.00	&0.00	&0.00	\\
PA. CR.			&0.10	&0.40	&0.40	&0.00	&0.00	\\
FI. TR.			&0.40	&0.00	&0.40	&0.00	&0.20	\\
\hline
\end{tabular}
\end{center}
\normalsize
The SIP responses are categorized according to their different classes. We set the CPT matrix of the \textsf{Response Distribution} to the next:
\footnotesize
\begin{center}
\begin{tabular}[htbc]{|l|c|c|c|c|c|c|}
\hline
Request			&1xx	&2xx	&3xx	&4xx	&5xx	&6xx	\\
\hline
NORMAL			&0.30	&0.30	&0.05	&0.05	&0.05	&0.05	\\
SCAN			&0.10	&0.05	&0.05	&0.70	&0.10	&0.00	\\
SPIT			&0.30	&0.20	&0.05	&0.20	&0.20	&0.05	\\
DOS			&0.20	&0.10	&0.20	&0.20	&0.20	&0.10	\\
PA. CR.			&0.20	&0.00	&0.10	&0.60	&0.05	&0.05	\\
FI. TR.			&0.30	&0.20	&0.05	&0.20	&0.20	&0.05	\\
\hline
\end{tabular}
\end{center}
\normalsize
Let us now fix the values of different variables observed at the child nodes and then assign the likelihood probabilities.
Let us assume that the attacker launches its queries into intervals of five seconds to give the appearance of a normal behavior. The calculus of the \textsf{Request Intensity} gives a value of $2.424$.
The result could be justified because such type of attack does not abnormally rise the \textsf{Request Intensity} contrary to what a DOS attack will do. 
The likelihood vector at the \textsf{Request Intensity} child is $\lambda = (1\ 0)$ because the intensity $=2.424 < 10$. The likelihood vector passed to the root node is $\lambda_{to\_parent} = (1\ 1\ 1\ 0\ 1\ 1)$. 
The calculus of the \textsf{ERI} gives a value of $2.70$.
The likelihood vector is $\lambda = (1\ 0)$ because the intensity $=2.70 < 4$. The likelihood vector passed to the root is $\lambda_{to\_parent}=(1\ 0.2\ 0.2\ 0\ 0\ 1)$. 
We assume that the \textsf{Number of Destinations} is 9 ($\lambda =(0\ 1)$ and $\lambda_{to\_parent}=(0\ 1\ 1\ 0.2\ 0\ 0.2)$) and that the \textsf{Number of RTP ports opened} is less than 10 ($\lambda=(1\ 0)$ and $\lambda_{to\_parent}=(1\ 1\ 0.8\ 0.8\ 1\ 0)$).
We assume that each dialog is closed before the subsequent dialog starts. The \textsf{Max Number of Dialog in Waiting State} is 1. $\lambda=(1\ 0)$ and $\lambda_{to\_parent}=(1\ 0.8\ 1\ 0.1\ 0.8\ 0.8)$.
The \textsf{Request Distribution} of the trace according to the SIP methods involved is next:
\begin{center}
\footnotesize
\begin{tabular}[htbc]{|l|c|c|c|c|c|}
\hline
method		&I	&R	&A	&C	&B \\
\hline
$\lambda$	&9	&0	&9	&1	&1  \\
\hline
\end{tabular}
\end{center}
\normalsize
$\lambda_{to\_parent}=(5.6\ 7.35\ 7.4\ 8.1\ 4.5\ 7.4)$.
The \textsf{Response Distribution} of the trace according to the response classes is next:
\begin{center}
\footnotesize
\begin{tabular}[htbc]{|l|c|c|c|c|c|c|}
\hline
class		&1xx	&2xx	&3xx	&4xx	&5xx	&6xx \\
\hline
$\lambda$	&7	&2	&0	&6	&2  	&0    \\
\hline
\end{tabular}
\end{center}
\normalsize
$\lambda_{to\_parent}=(3.1\ 5.2\ 4.1\ 3.2\ 5.1\ 4.1)$.

Now we can fuse all the $\lambda_{to\_parent}$ from children by elementwise multiplication: $L_i=\prod_c \lambda\_{to\_parent}_i (c);  1\leq i\leq 6$. 
We obtain the vector $L = (0\ 6.11\ 4.85\ 0\ 0\ 0)$. The belief about the trace is obtained by normalizing $L$ to the unit sum. $BEL=  (0\ 0.56\ 0.44\ 0\ 0\ 0)$. The trace is either \textsf{SCAN} or \textsf{SPIT}. This result could be explained because these two types of attacks are very similar. To refine the scales of different observed variables (intensities and high water marks) by using a greater number of intervals would better differentiate between these two types. 
\section{Related Works}
\label {related}
Intrusion detection systems (IDSs) have been deployed as a second defense line behind the intrusion prevention techniques and many research papers have been written on this topics. The conceptual building blocks of our method is the article of Valdes and Skinner in \cite{val:sri}.

 With the deployment of web services, the designers of IDSs recognize more and more the importance of a specific service knowledge such as implementation vulnerabilities and protocol weaknesses. A service specific anomaly detector which checks DNS and HTTP traffic is proposed in \cite{kru:service}. A multi-model approach to the detection of web based attacks has been recently published in \cite{kru:multimodal}. 

VoIP security became a major research topic over the last years. A general introduction to it can be found in \cite{jam:secu}. In \cite{pet:vul} we find a discussion about the vulnerabilities in small and home VoIP gateways and a proposal of some practical recommendations. Sensors at multiple layers to protect IP telephony from DOS attacks are inspired from works on TCP in \cite{rey:dos}. The authors of \cite{don:voi} highlights the difference between the control of the SPAM email and that of SPIT. They propose a voice SPAM control algorithm called Progressive Multi Gray-Leveling that fits in VoIP settings. In another way, the authors of \cite{ser:reb} propose social networks and reputation rating to deal with SIP SPAM.  More similar to our work is the proposed intrusion detection framework described in \cite{scidive}. Our approach is different with respect to it, since we use a Bayesian model capturing temporal and behavioral anomalies. 
\section{Conclusion}
\label{conclusion}
Intrusion detection and prevention is of major importance in VoIP networks. We have proposed in this paper an approach for VoIP intrusion detection based on prior work done in the intrusion detection community. We have justified the security needs in the VoIP environmet by a brief survey of the most relevant threads. The essential contribution is the modelling of SIP traffic and threat related entities so that SPIT, enumeration and denial of service can be detected. Other attacks remain still an open issue: man in the middle and call hijacking require a different kind of security mechanisms. A complementary solution is to avoid them, by managing the authentication and access control infrastructure. 
Future work will consist in releasing an open source based intrusion detection tool as well as performing real world operational evaluation of our solution.
\bibliographystyle{plain} 
\bibliography{paper1}
\footnotetext[1]{URLs are last explored on Feb 27, 2006}

\balancecolumns 
\end{document}